\documentclass[aps]{ws-procs9x6}

\usepackage{graphicx}
\usepackage{dcolumn}
\usepackage{amssymb}
\usepackage{bm}

\usepackage{enumerate}

\def\be{\begin{enumerate}}
\def\ee{\end{enumerate}}
\def\beq{\begin{equation}}
\def\eeq{\end{equation}}
\def\bea{\begin{eqnarray}}
\def\eea{\end{eqnarray}}
\def\a{\alpha}
\def\b{\beta}
\def\g{\gamma}

\def\d{\delta}

\def\l{\lambda}

\def\t{\tau}

 \def\half{\textstyle{\frac{1}{2}}}
\def\3halfs{\textstyle{\frac{3}{2}}} \def\em{\it}

\def\nab{\nabla}

\def\ben{\begin{enumerate}}
\def\een{\end{enumerate}}
\def\bitem{\begin{itemize}}
\def\eitem{\end{itemize}}

\def\uu{\underline{u}}

\begin{document}

\title{Einstein-\AE ther Theory\footnote{\uppercase{B}ased
on a talk given by \uppercase{T}.~\uppercase{J}acobson
at the \uppercase{D}eserfest.}}

\author{CHRISTOPHER ELING$^\dagger$, TED JACOBSON$^{\dagger,\ddagger}$, and DAVID MATTINGLY$^\S$}

\vspace{5mm}
\address{$^\dagger$~Dept. of Physics,
University of Maryland,
College Park, MD 20742-4111, USA\\
$^\ddagger$~Institut d'Astrophysique de Paris,
98bis Bvd. Arago, 75014 Paris, France\\
$^\S$~Dept. of Physics, University of California, Davis,
California, 95616, USA}



\maketitle

\abstracts{We review the status of ``Einstein-{\AE}ther theory", a
generally covariant theory of gravity coupled to a dynamical, unit
timelike vector field that breaks local Lorentz symmetry. Aspects
of waves, stars, black holes, and cosmology are discussed,
together with theoretical and observational constraints. Open
questions are stressed.}

\section{Introduction}

Could there be an {\ae}ther after all and we have just not yet
noticed it? By an ``{\ae}ther" of course we do not mean to
suggest a mechanical medium whose deformations correspond to
electromagnetic fields, but rather a locally preferred state of
rest at each point of spacetime, determined by some hitherto
unknown physics. Such a frame would not be determined by a
circumstance such as the moon's gravitational tidal field, or the
thermal cosmic microwave background radiation, but rather would
be inherent and unavoidable. Considerations of quantum gravity
have in multiple ways led to this question, and it has also been
asked in the context of cosmology, where various puzzles hint
that perhaps something basic is missing in the standard
relativistic framework.

Lorentz symmetry violation by preferred frame effects
has been much studied in non-gravitational physics,
and is currently receiving attention as a
possible window on quantum gravity.\cite{LVinQG}
But what about gravity itself?
General relativity is based on local
Lorentz invariance, so if the latter
is violated what becomes of the
former?
It is hard to imagine, both philosophically and technically,
how we could possibly give up general covariance,
the deep symmetry finally grasped through Einstein's long struggle.
Thus the question that interests us here is whether a generally
covariant effective field theory with a preferred frame
could describe nature.\footnote{More general sorts of
Lorentz violation in the gravitational sector are examined
in Ref.\refcite{Kgrav}.}

The simplest description of such a frame would appear to be via a
scalar field $T$, a cosmic time function, which has been proposed
in various contexts.\cite{JackiwPi,kflation,ghostcond1,ghostcond2}
The gradient $T_{,a}$, if timelike, defines a preferred rest
frame, and one can envision dynamics that would force it to be
everywhere timelike. But while a scalar field is simplest, the
norm of the gradient $|T_{,a}|$ is ``extra information", which
has nothing to do with specifying a frame {\it per se} but rather
specifies the rate of a particular cosmic clock. It may be that
Nature provides such a clock; we just wish to point out that the
clock rate is extra information. Constraining the gradient to
have fixed norm is not a viable option since, as explained in
section \ref{sec:Maxwell}, this would lead inevitably to caustics
where $T_{;ab}$ diverges. Another noteworthy feature of using a
scalar is that, by construction, the 4-velocity of the preferred
frame is necessarily hypersurface-orthogonal, i.e.~orthogonal to
the surfaces of constant $T$. Again, perhaps this is the way
Nature works, but it is a presumption not inherent in the notion
of a local preferred frame determined by microphysics.

The alternative discussed in this paper is to describe the
preferred frame by a vector field constrained kinematically to be
timelike and of unit norm, which we call the {\it {\ae}ther
field} $u^a$. Such a field is specified by three independent
parameters at each point, and generally couples via covariant
derivatives, so the theory is far more complicated than that of a
scalar time function. It is instinctive to worry about ghost modes
given a vector field without gauge invariance. However the unit
constraint on the vector renders it an unfamiliar beast. All
variations of the vector are spacelike, since they connect two
points on the unit hyperboloid, so ghosts need not arise.

There is a sparse history of studies of unit vector fields
coupled to
gravity.\cite{Gasperini,KosSam,Clayton,Barbero,Bekenstein,CLim,Lim}
Here we focus on the particular approach and results in which we
have been
involved.\cite{JM,MJindy,JMmatter,ElingJ1,JMwaves,ElingJ2} We
begin with the action principle that defines the theory, and then
discuss a Maxwell-like special case, linearized waves, PPN
parameters, energy, stars and black holes, and cosmology.

\section{Einstein-{\ae}ther action principle}
In the spirit of effective field theory, we consider a derivative
expansion of the action for the metric $g_{ab}$ and {\ae}ther
$u^a$.  The most general action that is diffeomorphism-invariant
and quadratic in derivatives is
\begin{equation} \label{eq:action}
S=\frac{-1}{16\pi G }
\int \!d^4x \sqrt{-g} \Bigl(R+K^{ab}{}{}_{mn} \nabla_a u^m \nabla_b
u^n +\lambda (u^a u_a -1)\Bigr)
\end{equation}
where
\begin{equation}
    K^{ab}{}{}_{mn}=c_1 g^{ab} g_{mn} + c_2 \delta^a_m \delta^b_n
    + c_3 \delta^a_n \delta^b_m + c_4 u^a u^b g_{mn}.
    \label{K}
\end{equation}
The coefficients $c_{1,2,3,4}$ are dimensionless constants, $R$
is the Ricci scalar, and $\lambda$ is a Lagrange multiplier that
enforces the unit constraint. The metric signature is
$({+}{-}{-}{-})$, and units are chosen such that the speed of
light defined by the metric $g_{ab}$ is unity. The constant $G$
is related to the Newton constant $G_{\rm N}$ by a
$c_i$-dependent rescaling to be discussed below. Other than the
signature choice we use the conventions of Ref.~\refcite{wald}.
The possible term $R_{ab} u^a u^b$ is proportional to the
difference of the $c_2$ and $c_3$ terms via integration by parts,
hence has been omitted. We have also omitted any matter coupling
since we are interested here in the dynamics of the
metric-{\ae}ther sector in vacuum. Note that since the covariant
derivative of $u^a$ involves the Levi-Civita connection, which
involves first derivatives of the metric, the {\ae}ther part of
the action in effect contributes also to the metric kinetic
terms. We call the theory with this action {\it Einstein-{\ae}ther
theory}, and abbreviate using ``\AE-theory".

Another way to express the theory is using a tetrad $e^a_A$ rather
than the metric, where $A$ is a Lorentz index. Then the {\ae}ther
can be specified as $u^a= u^A e^a_A$, with a unit Lorentz 4-vector
$u^A$ satisfying the constraint $\eta_{AB}u^A u^B=1$, where
$\eta_{AB}$ is the fixed Minkowski metric. This decouples the
normalization condition on $u^A$ from the dynamical metric. The
Lagrange density is then of the form $K^{ab}_{AB} D_a u^A D_b
u^B$, where $D_a$ is the Lorentz-covariant derivative involving
the spin-connection $\omega_a^{CD}$, and $K^{ab}_{AB}$ is a linear
combination of the four terms $g^{ab}\eta_{AB}$, $e^a_A e^b_B$,
$e^a_B e^b_A$, and $u^Cu^De^a_Ce^b_D\eta_{AB}$. This theory has a
local Lorentz invariance, which can be used to set the components
of $u^A$ to $(1,0,0,0)$. That produces the form of the theory as
presented by Gasperini.\cite{Gasperini} One can also use a
Palatini formalism, in which the spin connection is treated as an
independent variable to be determined via its field equation. In
this case the spin connection has torsion, because of the coupling
to $u^A$. If the solution $\omega_a^{CD}(e,u)$ is substituted back
into the action, one returns to the tetrad form, but with
different coefficients for each of the four terms in
$K^{ab}_{AB}$. The relation between these coefficients and the
original ones has not yet been worked out.

Given a metric and a unit vector field, there is a one parameter
family of metrics that can be constructed (aside from simple
rescalings). When expressed in terms of a different metric in this
family, the action changes, but only insofar as the values of the
$c_i$ in (\ref{K}) are concerned. More precisely, consider a field
redefinition of the form
\beq {g'}_{ab}={g}_{ab}+(B-1) u_au_b, \qquad
u'^a=Cu^a, \eeq
with $C = [1 + (B-1)u^2]^{-1/2}$ (where $u_a:=g_{am}u^m$ and
$u^2=g_{mn}u^m u^n$). Lorentz signature of both $g_{ab}$ and
$g'_{ab}$ requires $B>0$. The coefficient $C$ is chosen such that
$g'_{ab}u'^a u'^b = g_{ab}u^a u^b$, so the unit constraint is
unchanged, hence in the action we can put simply $C=B^{-1/2}$. The
action (\ref{eq:action}) for $(g'_{ab},u'^a)$ takes the same form
as a functional of $(g_{ab},u^a)$, but with different values of
the constants $c_i$.  The general relation between the $c_i$ and
the $c_i'$ has recently been worked out by Foster\cite{BZredef}.
His results reveal, for example, that one can arrange for
$c_1+c_3=0$ by choosing $B=1-c_1'-c_3'$ (provided $(c_1'+c_3')<1$
for Lorentz signature). A special case previously worked out by
Barbero and Villase\~{n}or\cite{Barbero} shows that the
{\ae}-theory is equivalent via field redefinition to GR when the
parameters satisfy\footnote{This corrects our earlier statement of
the equivalence in Ref.~\refcite{ElingJ1}. We thank B.Z.~Foster
for pointing out this error.} $c_1+c_4=0$, $c_1 +c_2+c_3=0$, and
$c_3=\pm\sqrt{c_1(c_1-2)}$. Lorentz signature implies $c_1<0$.
Note that if one first makes a field redefinition such that
$c_1+c_3=0$, then the Barbero-Villase\~{n}or result reduces to the
statement that the theory is equivalent to GR only if all
coefficients vanish.  Hence with the $c_1+c_3=0$ description we
ensure that non-zero coefficients always represent true deviations
from GR.

The field equation from varying the
{\ae}ther in the action (\ref{eq:action})
takes the form
\beq
\nab_a J^a{}_m -c_4 \dot{u}_a \nab_m u^a= \lambda u_m
\label{aethereqn}
\eeq
where
\beq J^{a}{}_{m} := K^{ab}{}_{mn} \nabla_b u^n, \eeq
and $\dot{u}^n = u^b\nab_b u^n$.
The field equation from varying the
metric in the
action (\ref{eq:action}) together with a matter action
takes the form
\beq
G_{ab}=T^{(u)}_{ab} + 8\pi G  T^{\rm matter}_{ab},
\label{AEeqn}
\eeq
where the {\ae}ther stress tensor is given by\cite{ElingJ1}
\begin{eqnarray}
T^{(u)}_{ab}&=&
\nab_m(J_{(a}{}^m u_{b)} - J^m{}_{(a} u_{b)} - J_{(ab)}u^m) \nonumber\\
&&+ c_1\, \left[(\nab_m u_a)(\nab^m u_b) -(\nab_a u_m)(\nab_b u^m)
\right]\nonumber\\
&&+ c_4\, \dot{u}_a\dot{u}_b\nonumber\\
&&+\left[u_n(\nab_m J^{mn})-c_4\dot{u}^2\right]u_a u_b \nonumber\\
&&-\frac{1}{2} g_{ab}L_u. \label{aetherT}
\end{eqnarray}
The constraint has been used in (\ref{aetherT}) to eliminate the
term that arises from varying $\sqrt{-g}$ in the constraint term
in (\ref{eq:action}), and $\lambda$ has been eliminated by
solving for it via the contraction of the {\ae}ther field equation
(\ref{aethereqn})
with $u^a$. The notation $L_u=-K^{ab}{}{}_{mn} \nabla_a u^m
\nabla_b u^n$ is the {\ae}ther lagrangian.

Our goal is to determine the theoretical and observational
constraints on the parameters $c_i$, and to identify phenomena
whose observation could reveal the existence of the {\ae}ther
field. For such phenomena one can look at post-Newtonian effects,
gravitational and {\ae}ther waves, and cosmology.

\section{Maxwell-like simplified theory}
\label{sec:Maxwell}

Before considering the general, rather complicated, theory it
makes good sense to ask if there is a simplification that might
serve at least as a decent starting point. A great simplification
occurs with the choice $c_1+c_3=0$ and $c_2=c_4=0$, so that the
connections drop out of the {\ae}ther terms in (\ref{eq:action}).
The {\ae}ther part of the Lagrange density then reduces to
\beq
2c_1u_{[a,b]}u^{[a,b]} + \l(u^2-1).
\label{EMax}
\eeq
This theory was studied long ago by
Nambu~\cite{Nambu} in a flat space context.
It is almost equivalent to Einstein-Maxwell theory in a
gauge with $u^2=1$. The difference is that one equation
is missing, since the action need only be stationary
under those variations of $u$ that preserve $u^2=1$.
The missing equation is an initial value constraint
equation, Gauss' law. If the current to which $u^a$ is
coupled is conserved, then Gauss' law holds at all
times if it holds at one time.

This theory coupled to dynamical gravity was first examined in
Ref.~\refcite{KosSam}, and further studied extensively in
Ref.~\refcite{JM} and Ref.~\refcite{Clayton}. In Ref.~\refcite{JM}
it was shown to be equivalent to Einstein-Maxwell theory coupled
to a charged dust, restricted to the sector in which there exists
a gauge such that the vector potential is proportional to the
4-velocity of the dust, i.e.~ the {\ae}ther field $u^a$. The
charge-to-mass ratio of the dust is given by\footnote{The
convention with Maxwell Lagrangian given by $-F^2/16\pi$ is
adopted here.} $(8\pi G/c_1)^{1/2}$. The extremal value
corresponds to $c_1=2$. A number of results were established
concerning static, spherically symmetric solutions, black holes,
and linearized solutions.

This case is appealing due to its simplicity, however a serious
flaw was noticed: solutions can have ``shocks" or caustics beyond
which the evolution of the {\ae}ther cannot be extended. In
particular\cite{JM}, consider the {\ae}ther configurations that
can be written as the gradient of a scalar $u_a=T_{,a}$, so that
the Maxwell-like ``field strength" tensor $u_{[a,b]}$ vanishes and
$u^a$ is orthogonal to the surfaces of constant $T$. Then the
field equations reduce to the vacuum Einstein equation, together
with the vanishing of the Lagrange multiplier $\l$ and the
statement that the gradient of $T$ is a unit vector, $T_{,a}
T^{,a}=1$. Then $u^a$ is necessarily a geodesic: \beq u^a
\nabla_a u_b = u^a \nabla_b u_a = 0. \eeq The first equality
holds since $u_b$ is a gradient, and the second holds since it is
a unit vector. If we launch geodesics orthogonal to an initial
surface of constant $T$, they will generically cross after some
finite proper time. Where they cross there is no well-defined
value of $u^a$, and the derivative $\nabla_a u_b$ is singular.
These are the shock discontinuities. A different demonstration of
the existence of shocks appears in Ref.~\refcite{Clayton}.

We note in passing that the preceding demonstration of shocks
applies in a very different context, namely the version of
$k$-essence~\cite{kflation} recently called ``ghost
condensation"~\cite{ghostcond1,ghostcond2}.\footnote{For a
discussion of caustics in a more general $k$-essence scenario see
Ref.~\refcite{kcaustics}.} This is the theory of a scalar field
$\phi$ with Lagrangian density of the form $P(X)$, where
$X=\phi_{,a}\phi^{,a}$, where $P(X)$ has a minimum at some value
$c$. Among the solutions to the field equations is the special
class which have $X=c$ everywhere. The above argument shows that
generically such configurations have caustics where $\phi_{;ab}$
is singular. In the cosmological setting, Hubble friction drives
all solutions to the minimum $X=c$. As far as we know it is an
open question whether the diverging gradients of the $X=c$
solutions is reflected in the generic cosmological solution.

Another flaw with the Maxwell-like case is that it admits
negative energy configurations, as shown by
Clayton~\cite{Clayton} using the Hamiltonian formalism in the
decoupling limit where gravity is neglected.\footnote{The
argument in Ref.~\refcite{Clayton} has a minor flaw, but the
conclusion is correct. The negative energy configuration
described there is a time-independent pure gradient
$u_i=\partial_i \phi(\vec{x})$. This initial data with vanishing
time derivative $u_{i,t}=0$ indeed has negative energy, however
the equation of motion implies that the time derivative does not
remain zero (unless $u_i=0$).} The existence of negative energy
configurations in this case is related to the fact that the
Lagrange multiplier $\lambda$ can be negative, so in the charged
dust interpretation the mass density is negative.

Before returning to the general class of Lagrangians
we note that one might also consider the theory where the
restriction on the norm of $u^a$ is enforced not rigidly
by a constraint but rather by a potential energy term
$V(u^a u_a)$ in the action. This approach was
discussed by Kosteleck\'{y} and Samuel~\cite{KosSam},
and more recently explored by Bjorken\cite{Bjorken:2001pe},
Moffat~\cite{Moffat:2002nu}, and
Gripaios\cite{Gripaios:2004ms}.
It has an additional, massive, mode,
which should be checked for a possible wrong
sign of the kinetic energy.

\section{Waves}

The
spectrum of linearized waves is important
for several reasons.
First, it can be used to constrain the theory
a priori, by
rejecting values of the parameters
$c_i$ for which waves carry negative energy or
for which there are exponentially growing modes.
Second, wave phenomena can be used to place
observational constraints on the parameters,
using radiation from compact objects such as the binary pulsar,
as well as cosmological perturbations.

The spectrum of linearized waves around a flat spacetime
background was worked out for the general theory defined by the
action (\ref{eq:action}) in Ref.~\refcite{JMwaves}.\footnote{The
Maxwell-like special case was previously treated in
Ref.~\refcite{JM}, and the case with only $c_1$ non-zero was
treated in Ref.~\refcite{MJindy}. In Ref.~\refcite{Lim} the modes
were found in the small $c_i$ limit where the {\ae}ther decouples
from the metric ({\it cf.} section~\ref{decoupling}).} The wave
modes in a de Sitter background were found in Ref.~\refcite{Lim}
(for $c_4=0$), which also studied the metric perturbations in
inflation interacting with the vector as well as a scalar
inflaton.

Here we summarize the results for the modes around flat space.
Since the {\ae}ther has three degrees of freedom, the total
number of coupled metric-{\ae}ther modes is five. There are two
purely gravitational (spin-2) modes, two transverse {\ae}ther
(spin-1) modes in which the {\ae}ther vector wiggles perpendicular
to the propagation direction, and one longitudinal or ``trace"
(spin-0) mode. The waves all have a frequency that is
proportional to the wave vector. Hence they are ``massless" and
have fixed speeds. The speeds for the different types of modes
are all different, and each mode has a particular polarization
type.

Table~\ref{table:waves} gives the speeds and polarizations for
the spin-2, spin-1, and spin-0 modes, in that order. The metric
and {\ae}ther have been expanded as $g_{ab}=\eta_{ab}+h_{ab}$ and
$u^a=\uu^a + v^a$, where $\eta_{ab}$ is the Minkowski metric and
$\uu^a$ is the constant background value that has components
$(1,0,0,0)$ in the coordinate system adopted. The gauge
conditions $h_{0i}=0$ and $v_{i,i}=0$ are imposed, where $i$
stands for the spatial components. The propagation direction
corresponds to $i=3$, and $I=1,2$ labels the transverse
directions. The notation $c_{123}$ stands for $c_1+c_2+c_3$, etc, and $s$
is the wave speed.
\begin{table}[h]
\tbl{Wave mode speeds and polarizations.}
{\begin{tabular}{@{}ll@{}}
\hline
squared speed & polarization\\
\hline
$1/(1-c_{13})$ & $h_{12}$, $h_{11}=-h_{22}$\\[1ex]
\vspace{2mm}$(c_1-\half c_1^2+\half
c_3^2)/c_{14}(1-c_{13})$ &
$h_{I3}=[c_{13}/s(1-c_{13})]v_I $\\[1ex]
$c_{123}(2-c_{14})/c_{14}(1-c_{13})(2+c_{13}+3c_2)$
& $h_{00}=-2v_0$\\
& $h_{11}=h_{22}=-c_{14}v_0$\\
& $h_{33}=(2c_{14}/c_{123})(1+c_2)v_0$\\
\hline
\end{tabular}\label{table:waves} }
\end{table}

\subsection{Stability}

The squared speed refers to the squared ratio of frequency to
wave-vector, so if it is negative for real wave-vectors the
frequency is imaginary, implying the existence of exponentially
growing modes.\footnote{The factor $s$ in the polarization
$h_{I3}$ of the transverse {\ae}ther mode implies that when
$s^2<0$ there is a $\pi/2$ phase shift of the metric perturbation
relative to that of the {\ae}ther.} The requirement that no such
modes exist restricts the parameters of the theory.\footnote{This
is not yet an {\it observational} constraint since it has not
been shown that the growing modes are not stabilized before their
effects become apparent.} For $c_i$ small compared to unity this
requirement reduces to the conditions $c_1/c_{14}\ge0$ for the
transverse vector-metric modes and $c_{123}/c_{14}\ge0$ for the
trace mode.

Lim argued\cite{Lim} that one should additionally demand that the
modes propagate subluminally (relative to the metric $g_{ab}$).
Although there is nothing wrong with local superluminal
propagation in a Lorentz-violating theory, he pointed out that
the vector field (in an inhomogeneous background) might tilt in
such a way as to allow energy on such locally superluminal paths
to flow around a closed spacetime curve. It is not clear to us
that it is really necessary to impose this extra demand, since
even in general relativity the {\it classical} field equations do
not forbid the formation of closed timelike curves, around which
relativistic fields could propagate. In any case, if we do make
this demand, then in the case of small $c_i$ it implies
$c_{13}\le0$, $c_1/c_{14}\le1$, and $c_{123}/c_{14}\le1$.

On the other hand, if gravitational waves propagate subluminally
relative to the ``speed of light" for matter, then matter can
emit gravitational \v{C}erenkov radiation. Using this phenomenon,
a very tight constraint on the difference between the maximum
speed of high energy cosmic rays and that of gravitational waves
was derived by Moore and Nelson~\cite{Moore:2001bv}. For cosmic
rays of galactic origin, the constraint is $\Delta
c/c<2\times10^{-15}$, while for extragalactic cosmic rays it is
$\Delta c/c<2\times10^{-19}$. \v{C}erenkov radiation in the
additional, {\ae}ther-metric modes has not been examined.
Constraints could conceivably eventually be obtained using these
processes.

The requirement that all the modes propagate on the light cone of
$g_{ab}$ is satisfied if and only if $c_4=0$, $c_3=-c_1$, and
$c_2= c_1/(1-2c_1)$ .

\subsection{Astrophysical radiation}

Since there are three additional modes, as well as
a modified speed for the usual gravitational waves,
one expects that the energy loss rate for orbiting
compact binaries will be affected. Not only are there
additional modes to carry energy, but also lower
multipole moments may act as sources, for example
there may well be a monopole moment generating the
spin-0, ``trace" mode. These phenomena
have not yet been studied.
Agreement with observations
of binary pulsars should yield constraints on the
$c_i$ parameters.

Another potential source of constraints is the
primordial perturbation spectrum in cosmology.
Lim\cite{Lim} has begun a study of this.

\section{Newtonian limit and PPN parameters}
\label{PPN}

Carroll and Lim\cite{CLim} have examined the Newtonian limit.
They adopt the ansatz of a static metric,
with the {\ae}ther vector parallel to the timelike Killing
vector. Restricting to the linearized field equations they
recover the Poisson equation $\nabla^2 U_N = 4\pi G_{\rm N}
\rho_{\rm m}$ for the gravitational potential $U_N$, where
$\rho_{\rm m}$ is the usual matter energy density and
\beq
G_{\rm N} = \frac{G}{1-c_{14}/2},
\label{GNewton}
\eeq
where $G$ is the
parameter appearing in the action (\ref{eq:action}).
Actually $c_4$ was set to zero in Ref.~\refcite{CLim},
but it can be restored without calculation by using
the fact\cite{ElingJ1} that in spherical symmetry
the effect of the $c_4$ term can be generated
by the replacements
$c_1\rightarrow c_{14}$ and $c_3\rightarrow c_3-c_4$.

The PPN formalism\cite{WillBook,WillReview} can be applied to
\AE-theory since it is a metric theory, at least in the
approximation (which is observationally known to be very
accurate) that the matter is minimally coupled in the usual way
to the metric.\footnote{Strictly speaking, if the matter couples
also to the {\ae}ther, producing local Lorentz violating effects
in the matter sector, then the PPN formalism must be modified to
allow for dependence of the PPN parameters on the composition of
the matter sources.} In the general setting there are ten PPN
parameters, but five of them vanish automatically in any theory
that is based on an invariant action principle, so we need
consider only the five remaining parameters,
$\beta,\gamma,\xi,\a_1,\a_2$.

The two PPN parameters $\beta$ and $\gamma$, known as the
Eddington-Robertson-Schiff (ERS) parameters,
are defined by the PPN expansion for the metric coefficients,
\bea
g_{00}&=&1-2U_N +2\b U_N^2+\cdots\label{PPN00}\\
g_{ij}&=&(1+2\g U_N +\cdots)\d_{ij}
\label{PPNij}
\eea
where $U_N$ is the Newtonian gravitational potential.
Thus $\beta$ controls the non-linearity and $\gamma$ the space
curvature due to gravity. In general relativity
$\b=\g=1$.
The field equations are apparently too complicated to
solve analytically, even assuming static, spherical
symmetry.
By numerical solution of the weak field equations in a
$1/r$ expansion it was found\cite{ElingJ1} that the metric takes
the form (\ref{PPN00},\ref{PPNij}), with $U_N$ is proportional
to $1/r$, consistent with the Newtonian limit.
Moreover, the ERS parameters take precisely the same values
as in general relativity in the generic case $c_{123}\ne0$.
(For the special cases with $c_{123}=0$ see Ref.~\refcite{ElingJ1}.)
Hence the theory is indistinguishable from GR at the
static, post-Newtonian level.

To expose the post-Newtonian differences from GR it is necessary
to examine the remaining PPN parameters, but these have not yet
been computed for the {\AE}-theory. The parameter $\xi$ is related
to preferred location effects, and likely vanishes in
{\AE}-theory. The parameters $\a_{1,2}$ are related to preferred
frame effects and almost surely do not vanish. They will
presumably arise when motion of the gravitating system relative to
the asymptotic {\ae}ther frame is allowed for.\footnote{A recent
preprint of Sudarsky and Zloshchastiev\cite{SZ} addresses this
point.}

One can hazard a guess based on the PPN parameters that were
calculated for the vector-tensor theory without the unit
constraint.\cite{WillBook} In that case, $\beta$ and $\gamma$ are
also unity in the case that corresponds most closely to
\AE-theory, namely $\omega=0$ in the vector-tensor parameters of
Ref.~\refcite{WillBook} together with $c_4=0$ in our
parameters\cite{ElingJ1}. If this agreement persists, one would
have in this case $\xi=\a_1=0$, but $\a_2\ne0$ generically.
However, in the special case when $c_{13}=0$, for which the
spin-2 waves propagate on the light cone of $g_{ab}$, the
vector-tensor parameters satisfy $\t=\eta$, and the result of
Ref.~\refcite{WillBook} yields $\a_2=0$. Thus perhaps in this
special case {\it all} the PPN parameters of Einstein-{\AE}ther
theory agree with those of GR.

Current limits on $\a_1$ and $\a_2$ are of order $10^{-4}$ and
$10^{-7}$ respectively, so constraints of this order on the
parameters of {\AE}-theory might be expected, at least for
generic parameters.

\section{Energy}

\subsection{Total energy}

As discussed in section~{PPN}, the Newtonian potential satisfies
the Poisson equation with source term $8\pi G_{\rm N}\rho_{\rm
m}$. This implies that in terms of the metric coefficient
$g_{00}\sim 1-r_0/r$, the source mass is given by $m=r_0/2G_{\rm
N}$. Assuming the source couples minimally to the metric
$g_{ab}$, this mass corresponds to an energy $m$ (since the speed
of light defined by $g_{ab}$ is one in our units). Accordingly,
one can infer that the energy of any isolated gravitating system
is given by the same formula. This differs from the ADM mass
$r_0/2G$ that one would have inferred from the action
(\ref{eq:action}). The reason is that the {\ae}ther stress tensor
(\ref{aetherT}) adds $1/r$ terms to the Newtonian field equation.

Another path to the same conclusion is to examine the energy via
its definition as the value of the Hamiltonian that generates
asymptotic time translations. Using Wald's Noether charge
formalism, Foster\cite{BZ} showed that when one takes into
account the falloff behavior of the fields at spatial infinity,
the {\ae}ther contribution to the energy flux integral takes the
form
\beq E^{(\rm {\ae}ther)}= -\frac{1}{8\pi G}\oint d^2S\;
K^{ta}{}_{rb}\, \nabla_a u^b, \label{Eaether} \eeq
where $K^{mn}{}_{ab}$ is the tensor defined in (\ref{K}).
An equivalent result was found by Eling\cite{Eling776}
using the Einstein pseudotensor.
It was shown in Ref.~\refcite{ElingJ1} that for a spherically
symmetric, static solution the line element has the
asymptotic form
\beq
ds^2= (1-\frac{r_0}{r}+\cdots)\, dt^2
- (1+\frac{r_0}{r}+\cdots)\, dr^2-r^2d\Omega^2
\eeq
and the {\ae}ther has a $t$-component of the form
\beq
u^t=(1-\frac{r_0}{2r}) + \cdots.
\eeq
The $r$-component of the {\ae}ther starts at $O(1/r^2)$, hence
does not contribute to (\ref{Eaether}). Using this form in
(\ref{Eaether}) one finds for the total energy
\beq
E=\frac{r_0}{2G}(1 - \frac{c_{14}}{2}).
\label{Etot}
\eeq
Using the relation (\ref{GNewton}) between $G$ and $G_{\rm N}$,
this can be re-expressed as $r_0/2G_{\rm N}$. Thus the total
energy is related to the ADM mass $M_{\rm ADM}=r_0/2G$ by a
constant rescaling.

\subsection{Positivity}

For coefficients $c_i$ such that $2 - c_{14}$ is positive,
positivity of the energy is thus equivalent to positivity of the
ADM energy. The usual positive energy theorem\cite{Witten} for GR
assumes the dominant energy condition holds for the matter stress
tensor, and proves that the total energy-momentum 4-vector of the
spacetime is future timelike. The {\ae}ther stress tensor
(\ref{aetherT}) does not appear to satisfy the dominant energy
condition (for any choice of the $c_i$), so the proof does not go
through as usual. Nevertheless, as discussed below, the energy of
the linearized theory is positive for certain ranges of $c_i$.
Perhaps a total divergence term that leaves the energy unchanged
must be removed before positivity can be seen. Also, since the
asymptotic value of the unit vector selects a preferred frame, it
might be that the energy is always positive only in that
particular frame. We can make no definite statement about the
non-linear energy at this time, based on general formal grounds.

\subsubsection{Linearized wave energy}
It is useful to examine the linearized theory to begin with. The
energy density of the various wave modes has been
found\cite{ElingJ2} using the Einstein or Weinberg pseudotensors,
averaging over oscillations to arrive at a constant average
energy density for each mode. The energy density for the
transverse traceless metric mode is always positive. For the
vector modes it is positive provided $(2c_1
-c_1^2+c_3^2)/(1-c_{13})>0$, and for the trace mode it is
positive provided $c_{14}(2-c_{14})>0$.  For small $c_i$ these
energy positivity conditions reduce to $c_1>0$ and $c_{14}>0$,
respectively.

In the Maxwell-like case $c_{13}=c_2=c_4=0$ the linearized energy
positivity requirement reduces to $0< c_1 < 2$. The negative
energy configurations discussed in section \ref{sec:Maxwell} do
not show up in the linearized limit.  Their energy density is
proportional to $-(\nabla u^0)^2$, which is quartic in the
perturbation $u^i$.

\subsubsection{Gravitational decoupling limit}
\label{decoupling}

The linearized waves are coupled metric-{\ae}ther modes. A
simpler limit to consider is a decoupling limit in which gravity
is turned off. To access this limit formally we can let $G$ and
$c_i$ tend to zero, while the ratios $c_i'=c_i/G$ are held fixed.
If the metric is expanded as $g=\eta+\sqrt{G} h$ in the action
(\ref{eq:action}), and the limit $G,c_i\rightarrow0$ is taken,
then one is left with just the action for linear gravitons and a
decoupled {\ae}ther action where all metrics are replaced by
$\eta$ and all covariant derivatives are replaced by ordinary
partial derivatives. This limit was studied by Lim\cite{Lim} in
the case $c_4=0$. Restoring the $c_4$ dependence one finds perfect
agreement between his results and the decoupling limit of the
coupled linearized waves.

As mentioned in section \ref{sec:Maxwell}, Clayton showed that
the energy can be negative in the Maxwell-like special case. He
also claimed that this remains true for more general choices of
the coefficients $c_i$. However, the case where only $c_1$ is
non-zero corresponds in the decoupling limit to a nonlinear sigma
model (NLSM) on the unit hyperboloid which, like all NLSM's, has
a stress tensor satisfying the dominant energy condition.

\subsection{Summary of constraints on the parameters}

So far we have discussed constraints from requirements in the
linearized theory of positive energy, stability (no exponentially
growing modes), and subluminal propagation (not necessarily
required). Taken together, the constraints of positive energy and
stability in the linearized regime imply $c_1>0$, $c_{14}>0$, and
$c_{123}>0$ (for small parameters). These are likely necessary
for a viable theory. The requirement of no superluminal
propagation (which we do not feel is necessary) would
additionally imply $c_{13}\le0$, $c_1\le c_{14}$, and $c_{123}\le
c_{14}$.

There is plenty of parameter space in which all the linearized
constraints one might think of imposing are satisfied. It remains
an important open question to determine whether energy positivity
can be ensured beyond the linearized limit. But one thing is
already notable, namely that this provides examples of a theory of
a vector field which has no standard gauge symmetry and yet which
has only stable, positive energy modes. The key factor making this
possible is the constraint on the norm of the vector.

\section{Stars and black holes?}

In static, spherical symmetry the {\ae}ther vector must be a
linear combination of the time-translation Killing vector and the
radial vector. Solutions were found in Ref.~\refcite{ElingJ1} for
which the {\ae}ther field has a radial component that falls off as
$1/r^2$. The question arises as to what happens to this vector in
the near field region of a star. Symmetry and regularity imply
that the radial component vanishes at the origin of spherical
symmetry, so if indeed a regular solution exists, $u^a$ must be
parallel to the Killing vector at the origin and at infinity but
not in between. Such solutions have recently been studied
perturbatively by Sudarsky and Zloshchastiev\cite{SZ}. We have
recently shown numerically that solutions with no radial component
of the {\ae}ther also exist\cite{ElingJ3}.

In the case of a black hole there is no regular origin
of spherical symmetry, but the question arises as to what
happens to $u^a$ on the horizon. The vector $u^a$ cannot
exist at a bifurcation surface (where the Killing vector
vanishes, like the 2-sphere at the origin $U=V=0$
of Kruskal coordinates).
The reason is that the Killing flow acts there as a Lorentz
boost in the tangent space of any point on the bifurcation
surface, hence would act non-trivially on $u^a$. Thus
$u^a$ could not be invariant under the Killing
flow. Since $u^a$ is constrained to be a unit vector
it cannot vanish, hence we infer that it must blow up as the
bifurcation surface is approached.

One might think that the impossibility of an invariant {\ae}ther
at the bifurcation surface implies there is no regular black hole
solution in this theory, since regularity on the future horizon
is somehow connected to regularity at the bifurcation surface. It
seems this is not necessarily the case. A result of R\'{a}cz and
Wald\cite{RaczWald} establishes, independent of any field
equations, conditions under which a stationary spacetime with a
regular Killing horizon can be globally extended to a spacetime
with a regular bifurcation surface, and conditions under which
matter fields invariant under the Killing symmetry can also be so
extended. In spherical symmetry the conditions on the metric are
met for a compact Killing horizon with constant, non-vanishing
surface gravity, so the result of Ref.~\refcite{RaczWald}
indicates that an extension to a regular bifurcation surface must
exist. However, one of the conditions on the matter (i.e.
{\ae}ther) field is not met, namely, it is not invariant under
the time reflection isometry. This is because the timelike vector
$u^a$ obviously breaks the local time reflection symmetry. Thus
the {\ae}ther vector need not be regular at the bifurcation
surface (although all invariants must be regular and, given the
Einstein equations, the {\ae}ther stress-tensor must remain
regular in the limit that the bifurcation surface is approached).
Hence, as far as we know, there is no argument forbidding regular
black hole solutions.

In fact, we have expanded the field equations about a regular,
static future event horizon and shown\cite{ElingJ3} that locally
regular solutions exist.
Using a shooting method, we have shown that
the free parameters can be chosen so that the solutions extend
to asymptotically
flat metrics at spatial infinity.
Alternatively, it
seems an attractive idea to numerically study the spherically
symmetric time-dependent collapse scenario. The collapsing matter
could be a scalar field, but more simply it could just be a
spherical {\ae}ther wave.

\section{Cosmology}

Finally we turn to the role of the {\ae}ther in cosmological
models. Assuming Robertson-Walker (RW) symmetry, $u^a$
necessarily coincides with the 4-velocity of the isotropic
observers, so it is entirely fixed by the metric. The {\ae}ther
field equation has but one non-trivial component, which simply
determines the Lagrange multiplier field $\l$. Therefore the
entire {\ae}ther stress tensor is also determined by the metric.
Like any matter field, when the {\ae}ther satisfies its equation
of motion, its stress tensor is automatically conserved. Hence,
in RW symmetry, the {\ae}ther stress tensor must be a conserved
tensor constructed entirely from the spacetime geometry. One such
tensor is the Einstein tensor itself. Another is the stress
tensor of a perfect fluid with equation of state
$p=-\frac13\rho$, whose energy density varies with the scale
factor $a$ as does the spatial curvature, i.e. as $1/a^2$. The
{\ae}ther stress tensor is just a certain combination of these two
conserved tensors\cite{MJindy,CLim}, namely
\beq T^{\rm {\ae}ther}_{ab}= -\frac{c_{13}+3c_2}{2}
\Bigl[G_{ab}-\frac16{}^{(3)}R(g_{ab}+2u_au_b)\Bigr]. \eeq
This is written using the conventions of
Refs.~\refcite{ElingJ1,JMwaves} in which the field
equations take the form (\ref{AEeqn}).

The effect of the cosmological {\ae}ther is thus to renormalize
the gravitational constant and to add a perfect fluid that
renormalizes the spatial curvature contribution to the field
equations. The renormalized, cosmological gravitational constant
is given by\cite{CLim}
\beq
G_{\rm cosmo}=\frac{G}{1+(c_{13}+3c_2)/2}.
\eeq
Carroll and Lim\cite{CLim} note that, since this
is not the same as $G_{\rm N}$, the expansion rate
of the universe differs from what would have been expected
in GR with the same matter content. The ratio is constrained
by the observed primordial ${}^4$He abundance to satisfy
$|G_{\rm cosmo}/G_{\rm N} - 1|<1/8$.
They assume the positive energy, stability, and subluminality
constraints discussed above, which imply
$G_{\rm cosmo}<G<G_{\rm N}$, so the universe would have
been expanding more slowly than in GR.
In our notation, the resulting helium abundance constraint
can be written as $15 c_1 + 21c_2+7c_3+8c_4<2$, where we have
included the $c_4$ dependence omitted in Ref.~\refcite{CLim}.


\section*{Dedication}

This paper is dedicated by TJ to Stanley Deser, with admiration
and gratitude for his friendship, support, and guidance in
gravitational exploration.

\section*{Acknowledgments}

We are grateful for helpful discussions during various stages of
this research with S.~Carroll, P.~Chru\'{s}ciel, T.~Damour,
J.~Dell, S.~Deser, G.~Esposito-Farese, A.~Kosteleck\'y,
K.~Kucha\v{r}, E.~Lim, C.~Misner, I.~R\'{a}cz, M.~Ryan,
A.~Strominger, M.~Volkov, R.~Wald, C.~Will, R.~Woodard, and
probably others to whom we apologize to for having omitted them
from this list.

This research was supported in part by the NSF under grants
PHY-9800967 and PHY-0300710 at the University of Maryland, by the
DOE under grant DE-F603-91ER40674 at UC Davis, and by the CNRS
at the Institut d'Astrophysique de Paris.

\end{document}